\documentclass{moriond}

\def\be{\begin{equation}}
\def\ee{\end{equation}}
\def\bea{\begin{eqnarray}}
\def\eea{\end{eqnarray}}

\usepackage{lipsum}
\usepackage{amsmath}
\usepackage{amssymb}
\usepackage{arydshln}
\usepackage{mathtools}
\usepackage{slashed}

\usepackage{colortbl}
\definecolor{Gray}{gray}{0.5}

\newcommand{\Photo}{}

\begin{document}
\vspace*{4cm}
\title{ HEAVY NEUTRAL LEPTONS BEYOND SIMPLIFIED SCENARIOS }

\author{\textbf{G. PIAZZA}$^{1}$\,\footnote{Speaker}, A. ABADA$^{1}$, P. ESCRIBANO$^{2\!}$, X. MARCANO$^{3}$  }

\address{$^{(1)}$IJCLab, Pole Th\'eorie, CNRS/IN2P3 et Universit\'e, Paris-Saclay, 91405 Orsay, France\\
$^{(2)}$ Instituto de F\'{\i}sica Corpuscular, CSIC-Universitat de Val\`{e}ncia, 46980 Paterna, Spain\\
$^{{(3)}}$ Departamento de F\'{\i}sica Te\'orica and Instituto de F\'{\i}sica Te\'orica UAM/CSIC,\\
Universidad Aut\'onoma de Madrid, Cantoblanco, 28049 Madrid, Spain
}

\maketitle\abstracts{
Heavy neutral leptons (HNL) constitute the building blocks of several neutrino mass generation mechanisms. Experimental searches depend on their masses and mixings with the active neutrinos, and exclusion regions in the plane of mass and mixing rely most of the time on two assumptions: $(i)$ the existence of $one$ HNL, which $(ii)$ mixes dominantly with only $one$ lepton flavor.
In this work we discuss how to reinterpret the limits from collider searches relaxing these assumptions, providing a simple recipe to recast the bounds in models with generic mixing patterns, and in which at least two HNLs are coupled to the active sector.
}

\section{Introduction}\label{Sec:Intro}
The generation of neutrino masses and mixings, as observed in neutrino oscillation phenomena, demands for an extension of the Standard Model (SM) that generally includes new fermions which, for masses above the MeV, are denoted as heavy neutral leptons (HNL). 
The presence of HNLs modifies the SM weak currents with a leptonic mixing matrix $U$, encoding the PMNS mixing matrix~\cite{Pontecorvo:1957cp,Maki:1962mu} and the active$-$sterile mixings. Depending on the mass regimes of the HNL, one expects impacts on numerous observables that give rise to a plethora of constraints (see Ref.~\cite{Abdullahi:2022jlv}
 and references therein). 

The relevant parameters entering in any process involving HNLs are their mass $M_N$ and mixings with the active sector ($U_{\alpha N}$, $\alpha=e,\mu,\tau$). Generally, the exclusion regions are given in the plane ($M_N, |U_{\alpha N}|^2$) and rely on the assumption of the existence of one HNL which mixes dominantly with only one lepton flavor. 
However, from neutrino oscillations experiments we know that {\it{at least}} two right$-$handed neutrinos are required in order to accommodate data. Moreover, bounds on the active$-$sterile mixings $(U_{eN},\,U_{\mu N},\,U_{\tau N})$ in scenarios with three or more HNLs are very relaxed \cite{Chrzaszcz:2019inj}\,, encouraging to push HNL searches beyond the simplified scenarios. It is the purpose of this work to show how the bounds on HNL from collider searches are recast in realistic setups 
with more than one HNL, and with general mixings patterns to the active flavors.

In the following we will remain agnostic about the mechanism of neutrino mass generation, considering the most generic case with masses and mixings as free parameters. 
In Section~\ref{Sec:CollSearch} we will briefly overview collider searches  in the mass regime where HNLs decay promptly into four leptons or two leptons and jets, while in Section~\ref{Sec:Recasting} we will describe the recasting procedure focussing on the leptons plus jets channel. We focus here on the $M_N < M_W$ regime, although it can also be applied to heavier masses. These results are based on Ref.\cite{PGAX}.

\section{Collider Searches}\label{Sec:CollSearch}
In the mass range $M_N \in [6, 80 ]\, \rm GeV$, HNL can be produced in $p$-$p$ collisions via on-shell W decay. 
 \noindent The relevant processes are (along with the charge conjugated channels)

\be
pp \to X + W^+ \to X + \ell_\alpha^+ + N \to \begin{cases}
X+ \ell_\alpha^+, \ell_\beta^\pm, \ell_\gamma^\mp, \nu_\gamma \\
X+ \ell_\alpha^+, \ell_\beta^\pm, n j 
\end{cases}, \quad  \alpha,\, \beta, \, \gamma = e,\, \mu,\, \tau
\label{eq:pp}
\ee
which can be Lepton Flavor Violating (LFV) and, if neutrinos are Majorana particles, Lepton Number Violating (LNV) processes.
 The bounds on ($M_N, |U_{\alpha N}|^2$) are extracted under the assumptions that a single HNL interacts mainly with one of the active flavors, namely $U_{	eN}\neq 0$ or  $U_{\mu N}\neq 0$.
 However, as described in Sec.~\ref{Sec:Intro}, realistic models require at least two HNLs to explain neutrinos oscillation data, with generic mixing patterns to the three families of active flavors.
  As a matter of fact, as shown in Ref.\cite{Tastet:2021vwp} for the $3 \ell + \slashed{E}$ searches at ATLAS\,\cite{ATLAS:2015gtp}, in the generic $3_{\rm active}+2_{\rm steriles}$ models (named $3+2$) bounds on HNLs parameters are strongly model dependent and can be relaxed by orders of magnitude in some specific implementations aimed to explain neutrino oscillation data.
 In the following, we will focus on the $2 \ell + n j $ channels, showing how the bounds extracted in the simplified scenarios can be significantly altered once recast in realistic models \cite{PGAX}. 

\section{Recasting the Bounds}\label{Sec:Recasting}
For $M_N < M_W$, the sterile neutrino is produced on-shell (narrow width approximation), thus the decay rate can be factorized as

\be
\Gamma(W^+ \to \ell_\alpha^+ \ell_\beta^\pm q \bar{q}') = \Gamma(W^+ \to \ell_\alpha^+ N)\,\times  \rm{Br}(N \to  \ell_\beta ^{\pm} q \bar q\prime)\, .\label{eq:Wdec}
\ee

\noindent The total rate scales as $  |U_{\alpha N}|^2  |U_{\beta N}|^2 / \Gamma_N^{\rm tot	} $, which in the single mixing approximation $\alpha=\beta$ reduces to $  |U_{\alpha N}|^2 $ since $\Gamma_N^{\rm tot	} \propto  |U_{\alpha N}|^2$.  Under these assumptions experimental collaborations put bounds on the single mixing $  |U_{\alpha N}|^2 $, with $\alpha = e$ or $\mu$, with the only exception of CMS \cite{CMS:2018jxx} that considered the combination $U_{e N}\,,U_{\mu N} \neq 0$.
However, in the generic 3+2 setup that we consider here, two effects can modify the $W$ decay rate: $(i)$ fixing the final states targeted by the experiment (generally $e^+ e^\pm$ and $\mu^+ \mu^\pm$), the total cross section will depend on all the active$-$sterile mixings through $\Gamma_N^{\rm tot	} $, and $(ii)$ if the two sterile neutrinos are almost degenerate in mass $\Delta M \ll {M}_{N_{1,2}}$ and have similar mixings to the active sector $U_{\ell {N_1}}\sim U_{\ell {N_2}}$, they can interfere producing a modulation effect depending on $\Delta M /\Gamma_N$ and the CP phases of the $U_{\ell {N}}$ matrix elements.
In the following, we will analyze the two cases separately.

\subsection{Generic Mixing Pattern (one HNL)}
 
In this case, fixing the active$-$sterile mixings in a given model, the total decay width of the sterile neutrino will be 

\begin{equation}
\Gamma_N^{\rm tot	} = |U_{eN}|^2\, \tilde \Gamma_{N}^e + |U_{\mu N}|^2\, \tilde \Gamma_{N}^\mu +|U_{\tau N}|^2\, \tilde \Gamma_{N}^\tau\,, 
\end{equation}
 
 \noindent with $\tilde \Gamma_{N}^\ell = \Gamma_{N}^\ell / |U_{\ell N}|^2$.
\noindent Thus, in order to recast the bounds on $ |U_{\ell N}|^2$ obtained in the single mixing approximation, one has to rescale it by

\be
 |U_{\ell N}|^2 \to  |U_{\ell N}|^2\ \times \frac{\Gamma_N^{\rm tot} }{\Gamma_N^{\rm sing} }\,.
 \label{eq:rescaling}
\ee
 with  $\Gamma_N^{\rm sing}= |U_{\ell N}|^2\, \tilde \Gamma_{N}^\ell $. 
 A straightforward generalization applies to the case where two mixings are considered simultaneously ($\Gamma_N^{\rm sing}\to  |U_{e N}|^2\, \tilde \Gamma_{N}^e  +  |U_{\mu N}|^2\, \tilde \Gamma_{N}^\mu$), like in the $W^+ \to e^+ \mu^\pm q \bar{q}'$ channel explored by CMS \cite{CMS:2018jxx}.
 
\noindent Notice that since $\Gamma_N^{\rm tot}  >\Gamma_N^{\rm sing} $, the bounds in the generic case will be less stringent.
\begin{figure}
\centerline{\includegraphics[width=.46\linewidth]{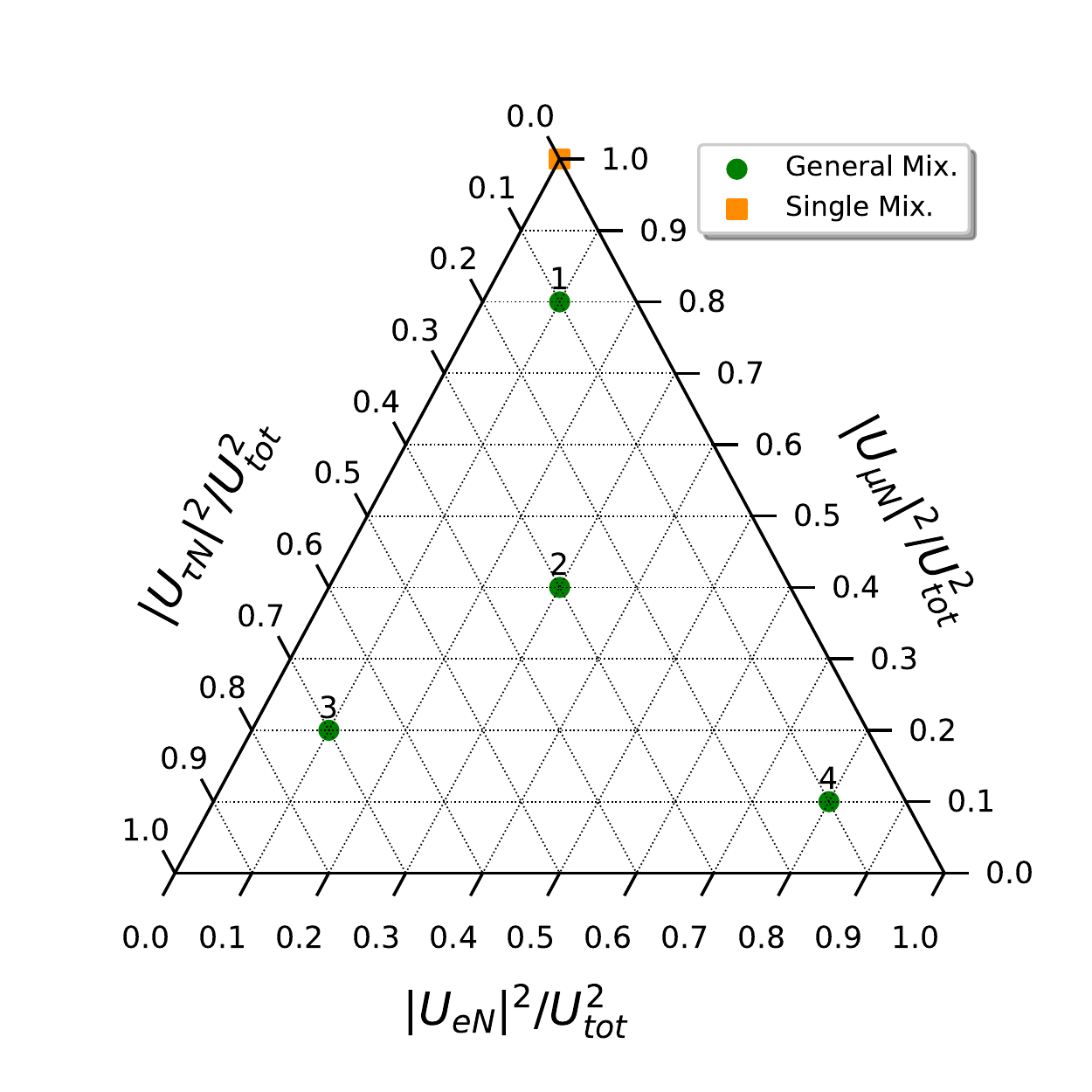}  \raisebox{3.4ex}{ {\includegraphics[ width=.58\linewidth]{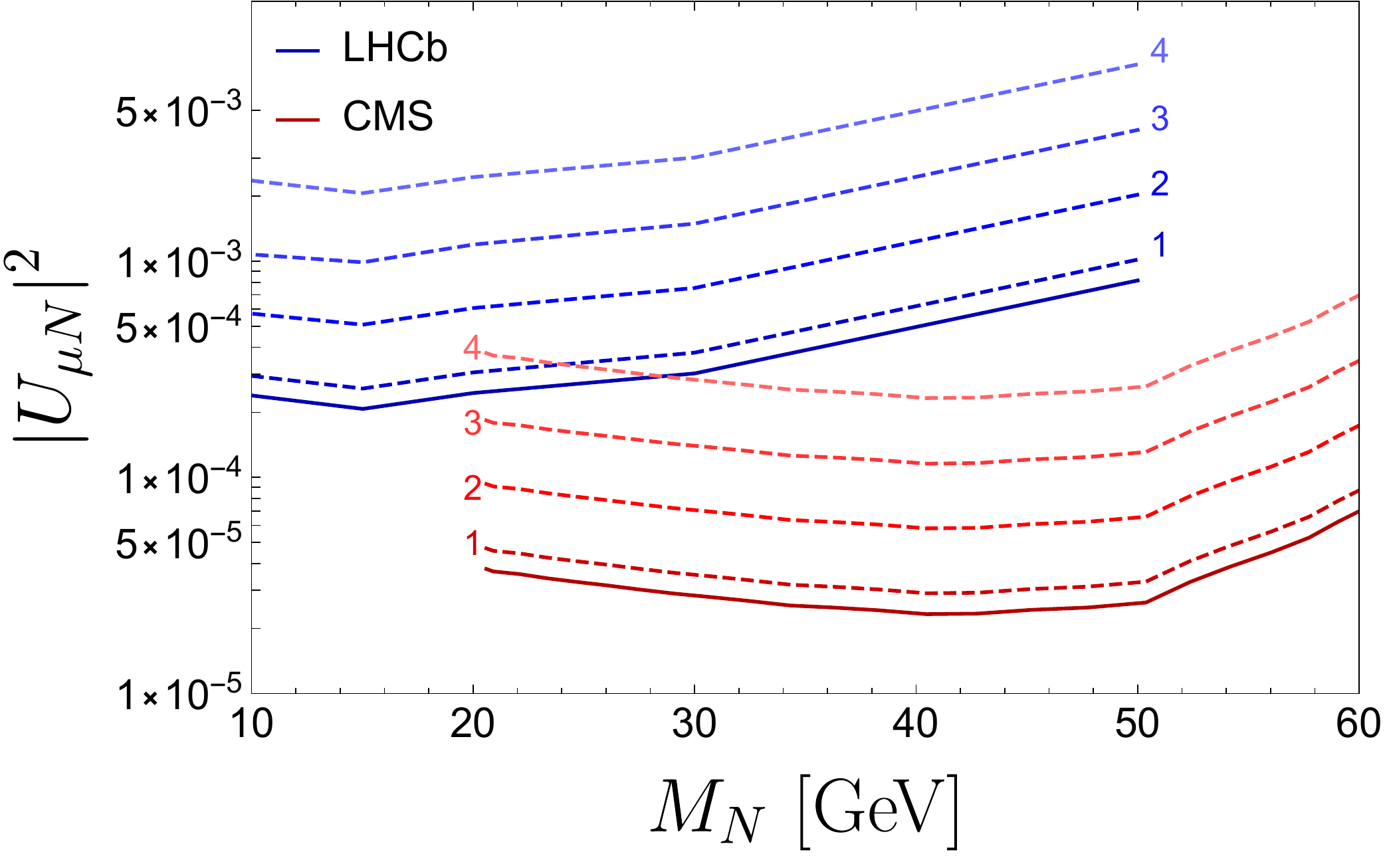} }}}
\caption[]{Rescaling of the bounds on $ |U_{\mu N}|^2$ from CMS\,\cite{CMS:2018jxx} (solid red line) and LHCb\,\cite{LHCb:2020wxx} (solid blue line), fixing the active$-$sterile mixings to the values corresponding to the green points $1-4$ in the ternary plot on the left. The orange squared point represents the single mixing case. }
\label{fig:ternary}
\end{figure}
\noindent To see the effect of the rescaling, we consider the bounds on $ |U_{\mu N}|^2$ from CMS\,\cite{CMS:2018jxx} and LHCb\,\cite{LHCb:2020wxx}, and appliy Eq.~(\ref{eq:rescaling}) for different combinations of the active$-$sterile mixings (Figure~\ref{fig:ternary}), with $U^2_{\rm tot}= |U_{e N}|^2+ |U_{\mu N}|^2+ |U_{\tau N}|^2$.
\noindent Moreover, we observe that unlike the case $3 \ell + \slashed{E}$, we do not need to reassess the signal efficiency as done in Ref.\cite{Tastet:2021vwp}, because in the generic mixing case no new processes can arise that contribute to the same final state $2 \ell +n j $.

\subsection{Two HNL's interference}
 
\begin{figure}
\centerline{\includegraphics[width=.75\linewidth]{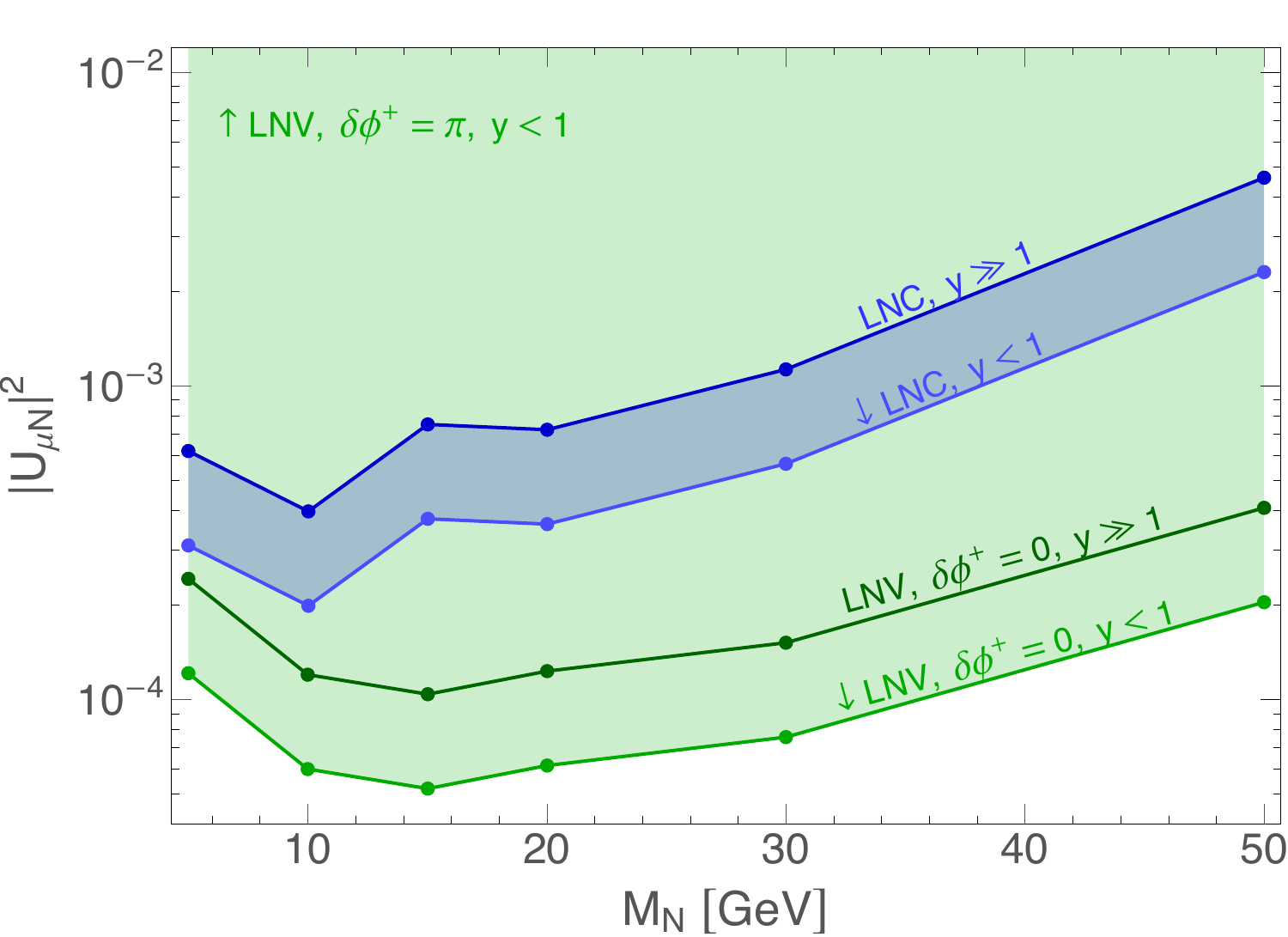}}
\caption[]{Rescaling of the bounds on $ |U_{\mu N}|^2$ from LHCb\,\cite{LHCb:2020wxx} in the presence of two HNLs. Dark blue line is the LHCb bound in the LNC searches, while in lighter blue (lower curve) is the rescaled bound for $y<1$. The dark green line is the bound in the LNV searches, which can be relaxed (upper green region) if the $N_{1,2}$ form a pseudo$-$Dirac pair ($\delta \phi^+=\pi$, $y\ll1$), or strengthened (lower green region) if $\delta \phi^+=0$.}
\label{fig:LHCb2NHL}
\end{figure}

 If the two sterile neutrinos $N_{1,2}$ are almost degenerate in mass, they contribute coherently to the same final state $2 \ell + n j $.
\noindent Assuming $M_{N_1} \simeq M_{N_2} \equiv M_N $, $\Gamma_{N_1} \simeq \Gamma_{N_2} \equiv \Gamma$ and $\Delta M\equiv M_{N_2}-M_{N_1} > 0$, the decay rate in the case of only one sterile neutrino factorizes as
\be \label{Gammas}
\Gamma(W^+ \to \ell_\alpha^+ \ell_\beta^\pm q \bar{q}')|_{N_1 \& N_2} = \Gamma(W^+ \to \ell_\alpha^+ \ell_\beta ^\pm q \bar{q}\prime ) |_{N_1} \times 2 \,\mathcal{K}(y, \delta\phi^\pm) \,,
\ee
and is modulated by a function $\mathcal{K}(y, \delta\phi^\pm) $, which in the limit $|U_{\ell N_1}|\approx |U_{\ell N_2}|$ is given by
\be
\mathcal{K}(y, \delta\phi^\pm) \equiv \left(1 + \cos{\delta \phi^\pm} \frac{1}{1+y^2} - \sin{\delta \phi^\pm}\frac{y}{1+y^2}\right)\,,
\ee
and where we have defined $U_{\ell_\alpha N_j}=  |U_{\ell_\alpha  N_j}| e^{i \phi_{\alpha j}}\,$, $\delta \phi^\pm = \big(\phi_{\alpha 2} - \phi_{\alpha 1} \big) \pm  \big(\phi_{\beta 2} - \phi_{\beta 1}\big)$ and $y= \Delta M /\Gamma$.
\noindent Therefore to recast the bounds we need to rescale the mixing to the flavor $\ell$ as
\be
 |U_{\ell N}|^2 \to  |U_{\ell N}|^2\ \times 2 \,\mathcal{K}(y, \delta\phi^\pm) \,.
 \label{eq:rescaling2}
\ee
Limiting to the case in which $\ell_\alpha=\ell_\beta$ (notice that $\delta \phi^-=0$), in Figure~\ref{fig:LHCb2NHL} we rescaled the bounds provided by LHCb\,\cite{LHCb:2020wxx}  for the LNV and LNC searches, for several values of $y$ and $ \delta\phi^+$.

\section{Conclusions}

We conclude that the bounds on HNLs in the simplified scenarios are often over-constraining if naively applied to realistic models explaining neutrino masses and mixings. The bounds are in fact model dependent and must be recast in order to be applied to such generic models.

\section*{Acknowledgments}
This project has received support from the European Union’s Horizon 2020 research and innovation programme under the Marie Skłodowska-Curie grant agreement No~860881-HIDDeN and from the Spanish Research Agency (Agencia Estatal de Investigaci\'on) through the Grant IFT Centro de Excelencia Severo Ochoa No CEX2020-001007-S and Grant PID2019-108892RB-I00 funded by MCIN/AEI/10.13039/501100011033.

\end{document}